# Effect of Number of Bilayers on the Anomalous Hall Effect in [Si/Fe]$_N$ Multilayers

S. S. Das and M. Senthil Kumar*

Department of Physics, Indian Institute of Technology Bombay, Mumbai 400076, India

The influence of varying the number of bilayers (N) on the anomalous Hall effect (AHE) in sputtered Si/Fe multilayers has been investigated. Both the AHE and magnetisation data reveal the in-plane magnetic anisotropy in the samples. Large enhancement of about 24 times in the saturation anomalous Hall resistance ($R_{hs}^A$) and anomalous Hall sensitivity (S) has been observed upon decreasing N from 20 to 1. When compared with the bulk Fe, the values of $R_{hs}^A$ and anomalous Hall coefficient, $R_s$ obtained for N= 1 were enhanced by about 5 and 3 orders of magnitude, respectively. The $R_s$ follows the longitudinal electrical resistivity $\rho$ as $R_s \alpha \rho^{2.1}$, suggesting side jump as the dominant mechanism of the AHE. The S as high as 22 Ω/T over a wide operational field range of -8 to +8 kOe has been obtained for N = 1.

## I. INTRODUCTION

Recently, multilayer structures of different combinations of materials are being studied extensively, not just due to their possible applications in the field of spintronics but also for the sake of understanding of magnetotransport properties at the fundamental level. Anomalous Hall effect (AHE), which is a spin-dependent transport phenomenon observed in magnetic materials, is still less understood despite being known for more than a century ago [1], [2]. Application of the effect in the field of magnetic sensing demands considerable enhancement in the sensitivity (S), anomalous Hall coefficient ($R_s$) of the materials. With the advancement of technologies, the research in the AHE field has been extended from homogeneous bulk magnetic systems to heterogeneous ferromagnetic nanostructures which includes layered structures (Fe/Cr, Fe/Cu, Fe/Gd, Co/Pd, Co/Pt etc.), metal-insulator nanocomposites ($FeSiO_2$, $CoSiO_2$ etc.) and thin films (Fe, Ni, Co etc.) [2]-[9]. More recently, Lu et al. have reported a high value of S = 12000 Ω/T in $SiO_2/FePt/SiO_2$ sandwich structure films [10]. Semiconducting materials like Ge, Si, GaAs, etc. used as an active layer in ordinary Hall sensors, have also been integrated with the ferromagnetic Fe layer so as to increase the applicability of the AHE-based sensors [11]-[13]. In our earlier paper on Si/Fe multilayers, we have studied the influence of both Fe and Si layers on the AHE [13]. A large enhancement (~60 times) of the AHE in this system has been observed by us. As a continuation of our research on AHE, in this paper, we present the effect of varying the number of bilayers N in the [Si/Fe]$_N$ multilayers.

## II. EXPERIMENTAL DETAILS

The [Si/Fe]$_N$ multilayers with different number of bilayers N were deposited onto glass and silicon substrates by DC magnetron sputtering at an argon pressure of 4 × 10$^{-3}$ mbar. The base pressure achieved prior to the deposition was 2 × 10$^{-6}$ mbar. The Si and Fe layer thicknesses of the multilayers were chosen as 50 and 20 Å, respectively. We have chosen these thicknesses as they yield large AHE as reported by us elsewhere [13]. The N of the multilayers was varied from 1 to 20. The structure and microstructure of the samples were studied through high-resolution transmission electron microscopy (HRTEM). The magnetic measurements both in in-plane and perpendicular directions were carried out using a vibrating sample magnetometer, an attachment of the physical properties measurement system (PPMS) of Quantum Design Inc.. The Hall effect measurements were performed at 300 K with the perpendicular magnetic field applied up to 27 kOe.

## III. RESULTS AND DISCUSSIONS

The anomalous Hall resistance ($R_h$) of the [Si(50 Å)/Fe(20 Å)]$_N$ multilayers measured at 300 K is plotted as a function of the applied field (H) in Fig. 1(a). The positive value of $R_h$ over positive field as seen in Fig 1(a) is an indication of hole-type carrier conduction in all the samples. The linear dependence of $R_h$ with H in the low field regime with negligible hysteresis is an indication of the magnetic anisotropy lying in the plane of the layers. For magnetic thin film samples, with an applied field perpendicular to the film plane, the $R_h$ can be expressed by an empirical relation [11], [14], [15],

$$R_h = \frac{\rho_h}{t} = \frac{R_0 H + R_s 4\pi M}{t} \qquad (1)$$

where $\rho_h$, $t$, $R_0$, $R_s$, and $M$ are Hall resistivity, thickness of the of the samples, the ordinary Hall coefficient, the anomalous Hall coefficient, and the perpendicular component of the magnetization, respectively. The first term of "1" corresponds to the ordinary Hall term arising due to the Lorentz force on the charge carriers because of the applied magnetic field and the second term represents the anomalous Hall term arising due to the breaking of right-left symmetry during the spin-orbit scattering of the electrons at disorder sites (impurities, surfaces, interfaces, etc.). The saturated anomalous Hall resistance ($R_{hs}^A$) can be estimated by extrapolating the high field regime Hall data to H = 0. We have determined the values of various Hall parameters, saturation anomalous Hall resistance ($R_{hs}^A$), $R_s$, $R_0$, and sensitivity (S) by following the procedure described in [3], [13], and [16].

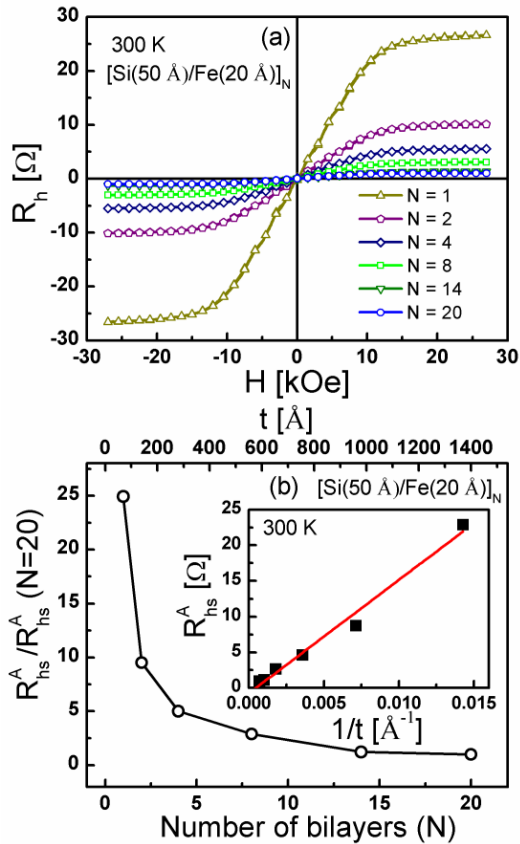

Fig. 1. (a) The $R_h$ versus H graph of the [Si(50 Å)/Fe(20 Å)]$_N$ multilayers of different N at 300 K. (b) The dependence of normalised saturation Hall resistance ($R_{hs}^A / R_{hs}^A(N=20)$) on N of the mulilayers. The open circles are the data points and the solid line is the guide to the eyes. The inset shows the 1/t dependence of $R_{hs}^A$.

From the analysis of the AHE data of the samples, the $R_{hs}^A$ obtained is normalized with respect to that for N = 20 and plotted as a function of N as in Fig. 1(b). A large enhancement of about 25 times in $R_{hs}^A$ has been observed upon decreasing N from 20 to 1. The largest value of $R_{hs}^A$ = 22 Ω (Fig. 1(a)) observed by us for N = 1 is about 5 orders of magnitude larger than that reported in the case of bulk Fe, i.e. 0.0002 Ω [6]. Similarly, the largest $R_s = 1.5 \times 10^{-7}$ Ω m/T obtained for N = 1 is about 3 orders of magnitude larger than that of bulk Fe, i.e. (2.8-7.2) $\times 10^{-10}$ Ω m/T [11], [15]. Such large enhancement in the AHE is often referred as "giant AHE" as an analogy to the "giant magnetoresistance" observed in magnetic multilayers. The AHE observed by us in the Si/Fe multilayers system is quite large (> $10^2$ times) when compared with the other multilayer systems with metallic spacer (Cu, Al, etc.) [13]. This suggests that the Si-Fe interface has an important role in the enhancement of the AHE. The interdiffusion between the ferromagnetic, Fe and semiconducting, Si layer results in the formation of a Si-Fe interfacial layer whose electrical nature is very complex due to the mixed metal-semiconducting nature of the charge carriers. The magnetic disorder that exists in this complex interface increases the scattering rate of the Hall carriers and hence the AHE. The short circuit and shunting effects [4] which are the main cause of the reduced AHE signal in metallic multilayers is significantly reduced upon the introduction of the Si spacer. Upon reducing the N of the multilayers from 20 to 1, the surface magnetic disorders of the samples increase, which results in the increase of the scattering of the charge carriers and hence an increase in the AHE signal. The $R_{hs}^A$ of the samples is plotted as a function of 1/t in the inset of Fig. 1(b) using the equation $R_{hs}^A = (R_s 4\pi M)/t$. The $R_{hs}^A$ of all the samples follows almost linear relationship with 1/t.

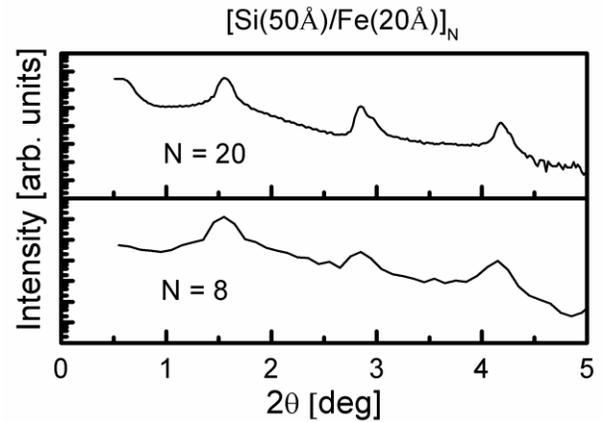

Fig. 2. The X-ray reflectivity data of the [Si(50 Å)/Fe(20 Å)]$_N$ multilayers samples for N = 8 and 20.

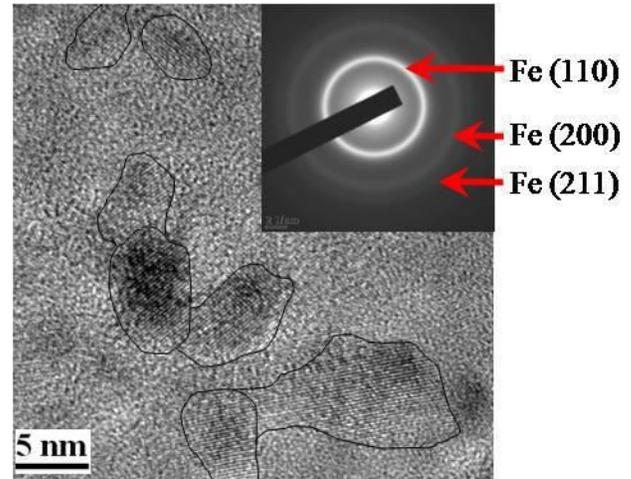

Fig. 3. The HRTEM image of [Si(50 Å)/Fe(20 Å)]$_2$ multilayers. The solid curves are drawn to show the grains. The inset shows the SAED pattern of the multilayer.

Since the AHE in the magnetic materials has its origin from magnetic scattering events, the microstructure and morphology of the individual ultra-thin layers, which can strongly influence the surface and interface effects in the Si/Fe multilayers worth a special mention. The XRD measurements on these samples show no peaks. This is consistent with the XRD results reported by us earlier, which showed no peaks for $t_{Fe}$ < 30 Å [16]. However, at relatively lower $t_{Si}$ values the existence of satellite Fe peak together with the iron-silicide peaks has been reported by Lucinski et al. [17] in [Fe(30 Å)/Si(14 Å)]$_{15}$ multilayers. The X-ray reflectivity (XRR) data (Fig. 2) taken on our samples show well-defined multilayer periodic structures for N ≥ 8. The XRR signal obtained for the samples with N < 8 is very weak to be detected by our instrument. Fig. 3 shows the representative HRTEM image and selected area electron diffraction (SAED) pattern (inset of Fig. 3) of the [Si(50 Å)/Fe(20 Å)]$_2$ multilayers. The SAED patterns of the samples show only the rings corresponding to Fe. The absence of Si rings confirms the amorphous nature of the Si layer. This is consistent with the results already reported by us [16]. The HRTEM images show nanocrystalline Fe grains. The HRTEM images taken on other samples also show similar features and no significant change in the microstructure of the samples with different N. The discontinuity of the Fe layer can be seen from the image, where the Fe grains (shown as solid curves) are widely separated from each other. Because of the discontinuity, even the grain boundaries of the Fe grains should be considered as surfaces/interfaces. Although there is no significant change in the microstructure (i.e. grain size and discontinuity) the enhancement of AHE indicates that the magnetic disorder at the interfaces/grain boundaries increases with decreasing N.

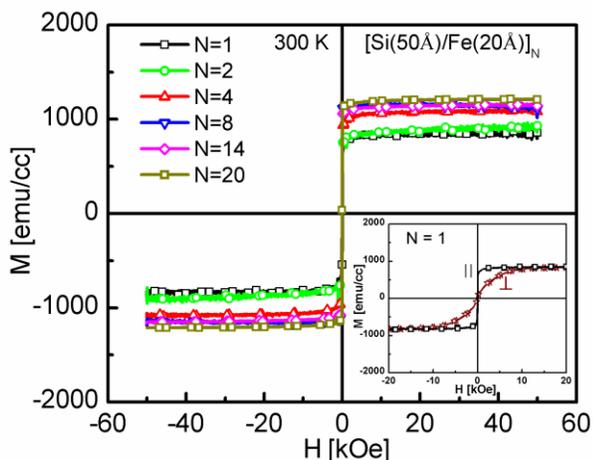

Fig. 4. In-plane magnetisation loops of the [Si(50 Å)/Fe(20 Å)]$_N$ multilayers at 300 K. The inset shows the representative in-plane and perpendicular MH loops for the sample with N = 1.

To understand further the AHE in the [Si/Fe]$_N$ multilayers, the magnetization measurements on the samples were performed at 300 K with the magnetic field applied both parallel and perpendicular to the film plane. Figure 4 shows the in-plane MH loops of the samples. The easy saturation of the in-plane MH loops with very small coercivity suggests that the samples are soft ferromagnetic, having in-plane magnetic anisotropy. The representative in-plane and perpendicular MH loops of the [Si(50 Å)/Fe(20 Å)]$_1$ sample as shown in the inset of Fig. 4 confirms the in-plane magnetic anisotropy in the sample. This is consistent with the AHE data of the samples. The saturation magnetization (M$_s$) of the samples remains almost constant around 1120 emu/cc for N ≥ 4 and shows a decrease around 860 emu/cc for the samples with N = 1 and 2.

This decrease of M$_s$ is normally observed in magnetic films of reduced thicknesses [13], [18]-[20]. Despite the decrease of M$_s$ for lower N, the AHE shows an enhancement. This suggests that the enhancement of the AHE due to magnetic disorder at the surfaces/interfaces is more dominant than the reduction due to the decrease of M$_s$.

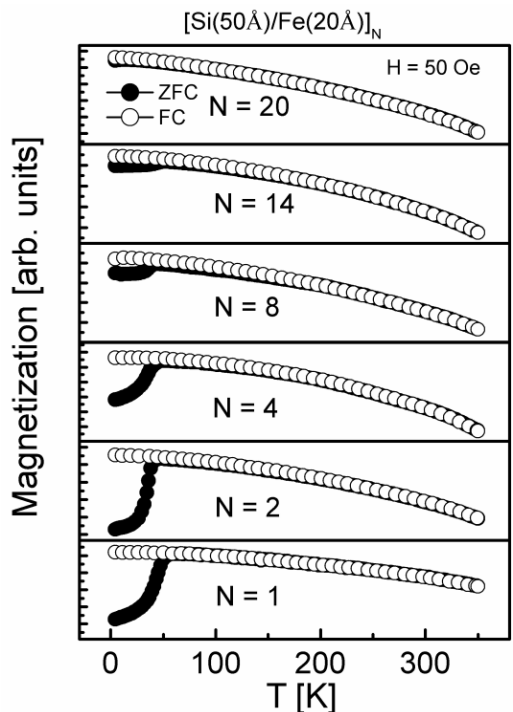

**Fig. 5.** The ZFC-FC curves of the [Si(50 Å)/Fe(20 Å)]$_N$ multilayers.

For further investigation of the magnetic properties, the temperature dependence of the magnetization of the samples were studied through zero field cooled (ZFC) and field cooled (FC) measurements. In the ZFC measurement, the samples were first cooled down to the lowest temperature of 4.2 K, then a small magnetic field of 50 Oe was applied parallel to the film plane and the data were taken while warming up the sample from 4.2 to 350 K. In the FC measurements, the samples were cooled to the lowest temperature 4.2 K in the presence of a low field of 50 Oe and the data were taken while heating up the sample from 4.2 to 350 K. Figure 5 shows the ZFC-FC curves of all the samples with N in the range 1-20.

As can be seen from Fig. 5, the ZFC and FC curves of the samples with N = 20 overlap in the entire temperature range. When N is reduced, the ZFC curves bifurcate from the FC curves at low temperatures. The amount of bifurcation increases with decreasing N. This behavior can be understood by considering the interlayer magnetic interaction between the neighboring Fe layers. Since we have fixed t$_{Fe}$ = 20 Å for all the samples, the same magnetic behavior of the Fe layers in all the samples with different N is expected. However, the observed behavior as seen in Fig. 5, is contrary to this. We presume that the small fraction of the nanograins in the Fe layers exhibits superparamagnetism, leading to the bifurcation for N = 1. As N increases, the interlayer exchange interaction between the neighboring Fe layers becomes stronger and stronger, thus stabilizing the magnetization of the nanograins. For N = 20, the interlayer interaction is so strong that the Fe layers are completely ferromagnetic and hence the bifurcation disappears. The maxima observed in the ZFC curves for N = 1 to 14 may be the superparamagnetic blocking temperature of the fraction of the Fe nanograins. These results indicate the coexistence of a small fraction of superparamagnetic grains with the ferromagnetic ones. Thus, the enhancement of AHE in our samples is due to the increase in magnetic disorder with decreasing N.

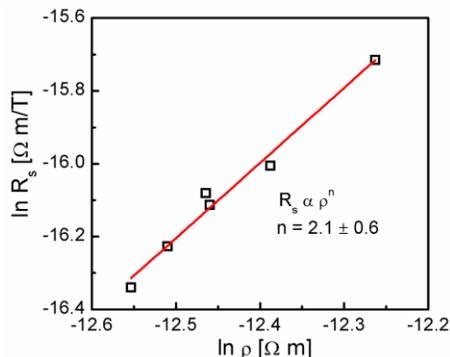

Fig. 6 The $\ln R_s$ versus $\ln \rho$ plot for the samples using the scaling law i.e. $R_s \propto \rho^n$. The open squares are the data points and the solid line is the fitted one.

To understand the mechanism of the AHE, we have tried to verify the scaling law for our Si/Fe multilayers. The scaling law has already been used for several other heterogeneous magnetic structures, like Fe/Gd, Fe/Cr, Co/Pd, Co/Pt, FeSiO$_2$, CoSiO$_2$, etc. [2]-[9]. Based on the scaling law, the $R_s$ and longitudinal electrical resistivity $\rho$ (H = 0) can be expressed empirically as, $R_s \propto \rho^n$ [13], [21], [22], where n = 1 and 2 correspond to the skew scattering and side jump mechanisms, respectively. It has been reported by many authors that skew scattering is dominant in low resistive materials, whereas in the case of high resistive materials, the side jump is the dominant mechanism of AHE [2]. Side jump as the dominant mechanism of the AHE has also been reported in the case of pure Fe [2]. The Linear fitting of the graph between $\ln R_s$ versus $\ln \rho$ as shown in Fig. 6, gives n = 2.1, suggesting side jump as the dominant mechanism of the AHE in the Si/Fe multilayers. Side jump as the dominant mechanism of the AHE has also been reported in the case of pure Fe [2].

The anomalous Hall sensitivity (S) obtained from the linear fitting of the low field regime Hall data shows an enhancement of about 24 times when N decreases from 20 to 1. The largest value of S = 22 $\Omega$/T was obtained for the sample with N = 1 is about three orders of magnitude larger than other multilayer systems with conducting spacers like Fe/Cu, Fe/Al, Fe/Cr, etc [13]. Our results show that the Si/Fe multilayers with sensitivity as high as 22 $\Omega$/T over a wide operational field range of -8 to +8 kOe could be a promising material in magnetic field sensing.

## IV. CONCLUSION

In conclusion, we have observed large enhancements in the AHE of the sputtered [Si/Fe]$_N$ multilayers upon decreasing N from 20 to 1. The largest value of $R_{hs}^A$ obtained for N = 1 is about five orders of magnitude larger than that of the bulk Fe. Similarly, the $R_s$ also shows an enhancement of three orders of magnitude when compared with the bulk Fe. Scaling law suggests side jump as the dominant mechanism of AHE. The large AHE observed in the Si/Fe multilayers envisages that the material could be a possible candidate for Hall sensors.